\documentclass{article} 

\usepackage[left=2.7cm,right=2.7cm,
top=1.6cm,                 
bottom=1.3cm,               
includehead,includefoot]{geometry}
\usepackage{amsmath}  
\usepackage{amsfonts} 
\usepackage{graphicx} 
\usepackage[utf8]{inputenc}  
\usepackage[colorlinks]{hyperref} 
\usepackage{amsmath,amssymb,amsthm,bm,bbm,amsfonts}

\begin{document}

\title{A first analysis of the ensemble of local maxima of maximal center gauge}

\author{Zeinab Dehghan\\
Department of Physics, University of Tehran, Tehran 1439955961, Iran
\\
zeinab.dehghan@ut.ac.ir
\and
Rudolf Golubich\\
Atominstitut, Technische Universit{\"{a}}t Wien, 1040 Wien, Austria
\\
rudolf.golubich@gmail.com
\and
Roman H\"ollwieser\\
Department of Physics, Bergische Universit{\"{a}}t Wuppertal, 42119 Wuppertal, Germany
\\
roman.hoellwieser@gmail.com
\and
Manfried Faber\\
Atominstitut, Technische Universit{\"{a}}t Wien, 1040 Wien, Austria
\\
faber@kph.tuwien.ac.at}


\date{\today}

\maketitle

\begin{abstract}
Maximal center gauge (MCG) aims to detect some of the most important vacuum configurations, suggesting thick magnetic flux tubes quantised to non-trivial center elements of the gauge group being responsible for confinement. Due to the NP-hardness of a global maximization of the gauge functional only numeric procedures aiming for local maxima are possible. We observe a linear decrease of the string tension with increasing gauge functional value of the local maxima. This implies that the request to get as close as possible to the absolute maximum is untainable. We compare global properties of the ensemble of local maxima with other methods to detect center vortices and with determinations of the string tension from full configurations. This comparison indicates that the information about the number and positions of center vortices is contained in the structure of the ensemble of local maxima. This may pave the way for a future more successful formulation of the gauge condition.
\end{abstract}

\section{Introduction}

The fundamental role of center vortices in the ground-state vacuum fields has been shown in Lattice QCD in the past decades~\cite{tHooft:1977nqb, tHooft:1979rtg, CORNWALL, DelDebbio:1996lih, Faber:1997rp, DelDebbio:1998luz, Bertle:1999tw, Faber:1999gu, Engelhardt:1999xw, Bertle:2000qv, Greensite:2003bk, Engelhardt:2003wm, Greensite:2016pfc}. The \textit{center vortex model} for quark-confinement assumes that closed quantised color-magnetic flux tubes percolating the QCD vacuum are responsible for a non-vanishing string tension. It  explains confinement, topological charge and chiral symmetry breaking~\cite{deForcrand:1999ms,Alexandrou:1999vx,Reinhardt:2000ck,Engelhardt:2002qs,Bornyakov:2007fz,hollwieser:2008tq,Hollwieser:2011uj,Schweigler:2012ae,Hollwieser:2012kb,Hollwieser:2013xja,Hoellwieser:2014isa,Hollwieser:2014mxa,Hollwieser:2014lxa,Greensite:2014gra,Hollwieser:2015koa,Trewartha:2015nna,Hollwieser:2015qea,Altarawneh:2015bya,Altarawneh:2016ped,Trewartha:2017ive,Faber:2017alm,Biddle:2018dtc} and reproduces the expected behavior of the string tension throughout a wide range of $\beta$-values, although the detection of center vortices in field configurations can proof difficult.

The main methods to detect center vortices are based on Maximal Center Gauge (MCG)~\cite{DelDebbio:1996lih,Langfeld:1997jx,Langfeld:2003ev} via maximizing the gauge functional $R_\text{MCG}$~(\ref{GaugeFunctR}) of the link variables. It is not possible to numerically identify a global maximum and one can only consider an ensemble of gauge copies $\{ \mathcal{C}_1, \mathcal{C}_2, \mathcal{C}_3, ..., \mathcal{C}_n \}$ which can be seen as a partially ordered set $ \mathcal{C}_1 \leq \mathcal{C}_2 \leq \mathcal{C}_3 \leq ... \leq \mathcal{C}_n $ with respect to $R_\text{MCG}$. During an optimization procedure the structure of the P-vortices can vary, but differences should decrease with the gauge functional approaching its maximal value. If they do not, one speaks of \textit{Gribov ambiguities} or \textit{Gribov problems} and care has to be taken when interpreting such gauge configurations. It turns out, however, that the P-vortices extracted by center projection from the gauge transformed configurations with maximal values of the gauge functional underestimates the string tension considerably.

We were able to mitigate this problem with the absolute maximum of the gauge functional by usage of non-trivial center regions \cite{Rudolf_Golubich_93027173,Rudolf_Golubich_89891800,Golubich_2021,Golubich_2020,Rudolf_Golubich_68347507,Rudolf_Golubich_65935846,golubich2022improvement}, where the gauge fixing is not performed solely on a microscopic level (taking only single lattice points and the links attached to them into account), but the non-trivial center regions, macroscopic structures guiding the gauge fixing procedure. This method is very involved and increasingly difficult for large lattices.

With the aim to circumvent the problem with the absolute maximum and to get a handier method for the identification of thick center vortices, we present an extended investigation of the ensembles of local maxima of the gauge functionals after a first test in~\cite{golubich93501217}. We create a reasonable large number of gauge copies and perform a gradient ascent towards the nearest local maximum of the gauge functional. After center projection, average values of Wilson loops $\langle W_\textrm{CP}\rangle$ and the quark-antiquark potentials are determined. This is different from Quantum Gauge Fixing~\cite{Langfeld_2000}, which does not only take the local maxima into account but suggests to integrate over the full gauge orbits with a gauge functional generating the weight function with an appropriately chosen scale. In SU(2) gauge theory we compare the results of ensemble averages to other center gauges such as \textit{adjoint Laplacian Landau gauge} (ALLG) and \textit{direct Laplacian Landau gauge} (DLCG) whose aim is to preselect appropriate, smooth gauge copies as starting points for a local maximisation of the gauge functional. The predictions of these methods we compare to the potentials measured on the unprojected (original) field configurations. The string tensions extracted from averages over the ensembles of local maxima give upper bounds of the string tensions. The individual values of the gauge functional demonstrate that in a wide range of gauge couplings the above mentioned extreme maxima do not contribute to the chosen large ensembles. The string tension extracted from the ensemble averages scales reasonably well with renormalization group predictions.

\section{Methods and Formalism}
We work on periodic lattices with $U_\mu(x) \in \text{SU(2)}$ denoting a gluonic link at lattice site $x$ pointing in direction $\mu$. We apply the Wilson action
\begin{equation}
S_\text{gluons} = \beta \sum_{x,\mu<\nu}\left( 1- \frac{1}{2}\mathfrak{Re}\text{Tr}(U_{\mu\nu}(x)) \right)
\end{equation}
with $U_{\mu\nu}(x)$ corresponding to a plaquette in the plane spanned by the directions $\mu$ and $\nu$ is used in a Monte Carlo procedure. For the inverse coupling $\beta = \frac{4}{g^2}$ we choose $\beta=2.3, 2.4, 2.5, 2.6, 2.7$. 

Center vortices are usually detected in \textit{direct maximal center gauge} (DMCG)  that uses a gradient ascent to identify the gauge transformation $g(x) \in \text{SU(2)}$ maximizing the gauge functional
\begin{equation}\label{GaugeFunctR}
R_\text{MCG}=\frac{1}{2N_l}\sum_x\sum_\mu\mid\text{Tr}[{}^gU_\mu(x)]\mid^2
\end{equation}
with ${}^gU_\mu(x)=g(x+\hat\mu)U_\mu(x)g^\dagger(x)$ and $N_l$ the number of links. Starting on multiple random gauge copies, this procedure is executed and the gauge copy with the largest value of the gauge functional is taken for further processing:  The link variables ${}^gU_{\mu}(x)$ are projected to the center degrees of freedom, that is $\pm1$ for SU(2),
\begin{equation}\label{CentrProj}
{}^gU_\mu(x)\rightarrow Z_\mu(x)\equiv \mathrm{sign Tr}[{}^gU_\mu(x)].
\end{equation}
The resulting non-trivial center projected plaquettes are known as \textit{P-plaquettes}, $U_\Box=-1$, which contain one or three non-trivial links. The duals of P-plaquettes form closed surfaces in the four dimensional dual lattice and the dual P-vortices correspond to the closed flux lines evolving in time. Center projection is the common ground of center detection methods, only the specific gauge fixing procedures vary. Since $\text{Tr}\,U\,\text{Tr}\,U^\dagger=\text{Tr}\,U_A+1$, where 
\begin{equation}\label{UA}
[U^A]_{ij} = \frac{1}{2}\text{Tr}[\sigma_i U\sigma_j U^\dagger]
\end{equation}
is the adjoint representation, MCG also maximizes
\begin{equation}\label{RAL}
R_\text{AL}=\sum_{x,\mu}\text{Tr}_A[{}^gU_\mu^A(x)],
\end{equation}
the gauge functional of the Landau gauge in the adjoint representation, which is blind to center elements. Landau gauge finds the configuration closest to a trivial field configuration. Hence, the functionals $R_\text{MCG}$ and $R_\text{AL}$ may fail to detect vortices in smooth field configurations, especially in the continuum limit. As was raised by Engelhardt and Reinhardt~\cite{Engelhardt:1999xw} in the context of continuum Yang-Mills theory a thin vortex gets singular in this limit and must fail to approach a smooth configuration.

Gauges not directly suffering from Gribov copies are Laplacian gauges which require to solve the lattice Laplacian eigenvalue problem. They were first applied in Laplacian Landau gauge by Vink and Wiese~\cite{Vink:1992ys} for the detection of Abelian monopoles and modified to Laplacian center gauge in \cite{Alexandrou:1999vx,deForcrand:2000pg} for center vortices which requires solving the eigenvalue problem of the lattice Laplacian in the adjoint representation
\begin{equation}\label{lapoperator} 
D_{ij}(x,y) = \sum_{\mu} 
\left(2\delta_{xy}\delta_{ij}-[U^A_\mu(x)]_{ij}\,\delta_{y,x+\hat{\mu}} 
-[U^A_\mu(x-\hat\mu)]_{ji}\,\delta_{y,x-\hat\mu}\right).
\end{equation} 
The three lowest eigenvectors of this Laplacian are combined to matrices $M(x)$ in \textit{adjoint Laplacian Landau gauge} (ALLG) which maximizes the functional
\begin{equation}\label{GaugeFunct}
R_\text{ALLG}= \sum_x\sum_\mu\text{Tr}[M^\dagger(x)U^A_\mu(x)M(x+\hat\mu)]
\end{equation}
with $M(x) \in SO(3)$ "on average", that is, with weakened orthogonality constrained $\langle M^\dagger \cdot M  \rangle = 1$. Even if the solution of the eigenvalue problem is a unique procedure, small modifications of the configuration may lead to different eigenfunctions corresponding to different positions of vortices. The combination of ALLG followed by DMCG~\cite{Faber_2001,Faber_2002} is referred to as \textit{direct Laplacian center gauge} (DLCG).

\subsection{The problem of the gauge functional}
To find the absolute maximum of the gauge functional~(\ref{GaugeFunctR}) is an NP-hard problem, hence it is practically impossible to find the absolute maximum. Moreover, the local numeric maximization of the gauge functional suffers from a severe problem as was discussed first by Engelhardt and Reinhardt~\cite{Engelhardt:1999xw} in the continuum and then numerically by Kovacs and Tomboulis~\cite{kovacs:1998xm} and by Bornyakov, Komarov and Polikarpov~\cite{Bornyakov_2001} with simulated annealing: The largest maxima of the functional  underestimate the density of vortices $\varrho_\text{vort}$. For smooth vortex surfaces the vortex density gives an upper limit for the string tension~\cite{Rudolf_Golubich_93027173}
\begin{equation}
\sigma \lessapprox  -\ln(1-2 \varrho_\text{vort}).
\end{equation}
Hence, a reduced vortex density comes with a reduced string tension for smooth vortex surfaces. We describe now a direct method to obtain the false local maxima, showing the origin of this problem.

\begin{figure}[!htb]  
	\centering
	\includegraphics[width=0.45\linewidth]{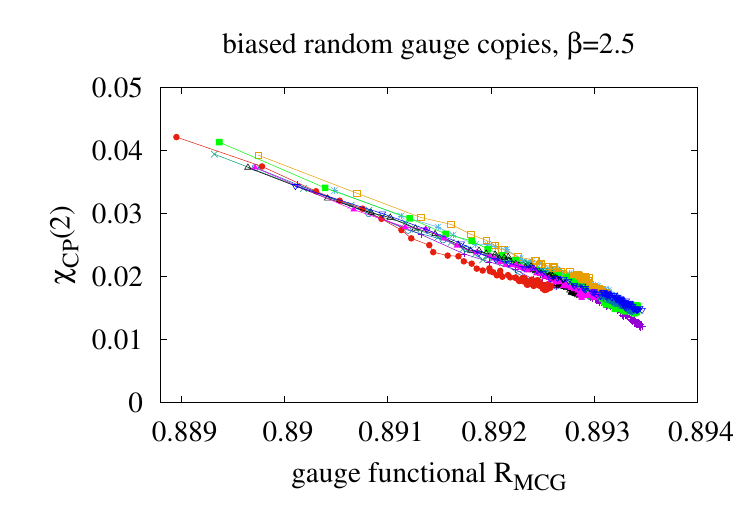}
	\includegraphics[width=0.45\linewidth]{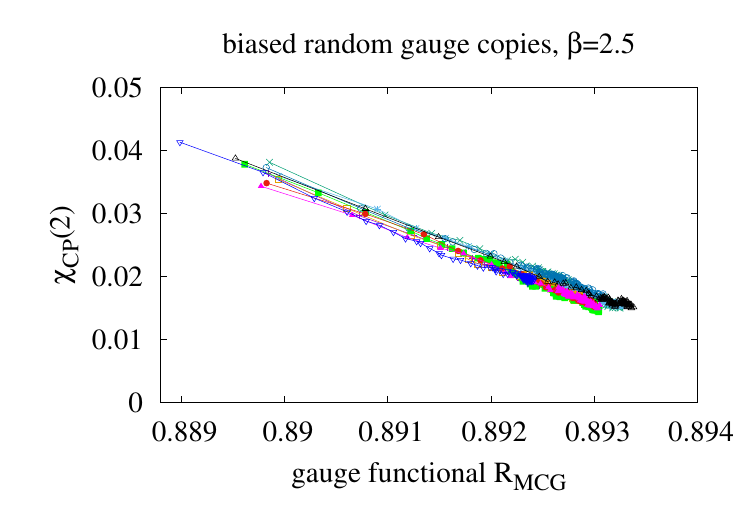}
	\caption{In the two diagrams we show the maximizing histories of two independent field configurations on a $24^4$ lattice for ten random gauge copies each. Interchanging hundred times gradient ascents and biased random gauge transformations, we get hundred points in the $\chi_\textrm{CP}-R_\text{MCG}$ plane connected by a line. Observe the nearly linear relation between Creutz ratio $\chi_\textrm{CP}$ and gauge functional $R_\text{MCG}$ and especially the final low value of $\chi_\textrm{CP}$, much lower than the expected string tension $\sigma=0.035$~\cite{Bali1994}.}
	\label{fig:biasedgauge}
\end{figure}

Random gauge copies are usually determined by random gauge matrices $g(x)$ with a homogeneous density on the SU(2)$\cong\mathbb S^3$ group manifold, as produced by the Box-Muller method~\cite{BoxMuller}. One can directly produce consecutively increasing local maxima of the gauge functional using a bias in the random gauge transformations by restricting them to the trivial sector of the residual $Z(N)$ gauge group which the maximal center gauge condition~(\ref{GaugeFunctR}) leaves free. For $N=2$ we have to choose random gauge matrices with positive trace. In the two diagrams in Fig.~\ref{fig:biasedgauge}, we show the maximizing histories for two independently created field configurations (out of 200) and for ten of their random gauge copies. For each of these 20 field configurations we perform 100 times the following two steps: First, we maximize the functional by gradient ascent. Secondly, we apply the biased random gauge transformations. For center projected field configurations of sufficient large lattice size already one configuration can reasonable estimate the string tension $\sigma$ from Creutz-ratios
\begin{equation}\label{Eq:Creutz}
\sigma\approx\chi_\textrm{CP}(R)=-\ln\frac{W_\textrm{CP}(R+1,R+1)\;W_\textrm{CP}(R,R)}{W_\textrm{CP}(R,R+1)\;W_\textrm{CP}(R+1,R)},
\end{equation}
for small Wilson loops $W_\textrm{CP}(R,T)$ of size $R \times T$. In Fig.~\ref{fig:biasedgauge} we show $\chi_\textrm{CP}-R_\text{MCG}$ pairs for the evolution through the biased gauge fixing iterations. A nearly linear relation between $\chi_\textrm{CP}$ and $R_\text{MCG}$ can be observed. With increasing gauge functional we observe a linear decrease of the string tension, underestimating the expected value of $\sigma=0.0350(12)$~\cite{Bali1994}.
With this method we did not find any configuration where the maximum of the gauge functional did not dramatically underbid the string tension.

\subsection{The importance of the ensemble}
DMCG predicts the string tension successfully~\cite{DelDebbio:1998luz,hollwieser:2008tq} when using a few gauge copies and choosing the one with the highest value of the functional for the further analysis, avoiding the global maximization problem. In DLCG~\cite{Faber_2001,Faber_2002} the problem is circumvented using the eigenfunctions of the adjoint Laplacian operator to select a smooth gauge field before maximizing the gauge functional in a gradient ascent. In this article, we use ensemble averages of maximal center gauge (EaMCG) to investigate the relation between the average value of the gauge functional and the average value of the string tension extracted from projected Wilson loops. We produce random gauge copies with the correct $\mathbb S^3$-weight, approach the next local maximum by the gradient method and take the average of an ensemble of such random gauge copies. The idea is that not the best local maximum alone carries the physical information, but the ensemble of local maxima.

Performing the gauge fixing procedure with a gradient ascend defines a map $f$ of gauge transformations $g$ to the gauge $g_\mathrm{max}$ corresponding to the nearest local maximum
\begin{equation}\label{Eq:gmax}
f: g \rightarrow g_\mathrm{max}
\end{equation}
Every $g_\mathrm{max}$ has a catchment area $\mathcal A_\mathcal{C}(g_\mathrm{max})$ such that
\begin{equation}\label{Eq:Einzug}
f(g)=g_\mathrm{max}\quad\textrm{for}\quad
g\in\mathcal A_\mathcal{C}(g_\mathrm{max}),
\end{equation}
serving as an implicit weight in a functional integral over gauge transformations. Every observable $O(\,^g\mathcal{U})$ is therefore mapped to $O(\,^{g_\mathrm{max}}\mathcal{U})$ via
\begin{equation}\label{Eq:Bahn}
\langle O(\mathcal{U})\rangle_S
=\int\mathcal DU\,\mathrm e^{-S(\mathcal U)}\int\mathcal Dg\, O(\,^{f(g)}\mathcal{U}).
\end{equation}
Maxima with an improbably high value of the gauge functional result in a reduced string tension, but they do not dominate the ensemble as our numerical results will show. The same holds for lower valued maxima, which overestimate the string tension.

\section{Numerical Results and Discussion}
In Monte-Carlo runs on a $24^4$ lattice we did for each $\beta\in\{2.3, 2.4, 2.5, 2.6, 2.7\}$ 20 random starts, 3000 initial sweeps and produced 1000 configurations with a distance of 100 sweeps. For these sets of $20\,000$ configurations we determined the averages of unprojected Wilson loops. Due to the large computational needs we did center projection by ALLG, DLCG and EaMCG for every 100th of these $20\,000$ configurations only. For EaMCG we did 100 unbiased gauge copies for each of the 200 configurations. In Fig.~\ref{fig:stringtension2} we compare the $24^4$ data for unprojected configurations with the same number of analogously produced $32^4$ configurations.

\subsection{Gauge functional distributions and Creutz ratios}
\begin{figure}[h!]
	\centering
	\includegraphics[width=\linewidth]{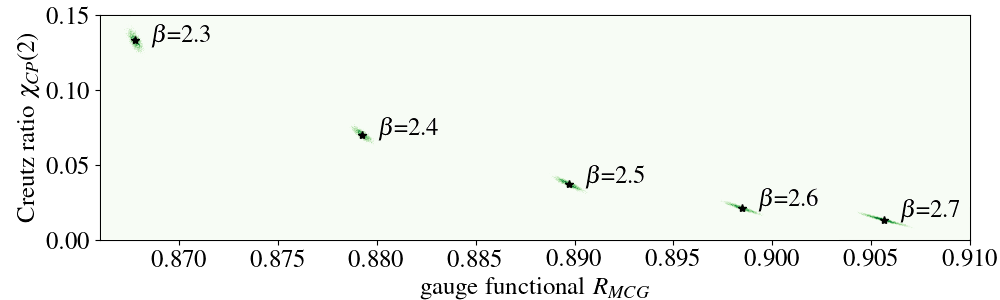}
	\caption{Correlations between Creutz ratios $\chi_\textrm{CP}(2)$ from center projected configurations and their values of the maximized gauge functionals $R_\text{MCG}$ for $\beta\in\{2.3, 2.4, 2.5, 2.6, 2.7\}$ on $24^4$-lattices. We produced density plots from 20 random starts and 10 configurations with a distance of 10000 MC iterations and 100 gauge copies each, the average values of the distributions are marked by a ``$\star$'' symbol.}
	\label{fig:densityPlots}
\end{figure}

In Fig.~\ref{fig:densityPlots} we show the distribution of 20\,000 gauge functional $R_\text{MCG}$ vs. Creutz ratio $\chi_\textrm{CP}(2)$ pairs for $\beta\in\{2.3, 2.4, 2.5, 2.6, 2.7\}$. For all $\beta$-values we see a clear linear correlation as expected from Fig.~\ref{fig:biasedgauge}. The average slope of the distributions agrees in both figures (not only for the case of $\beta=2.5$ shown in Fig.~\ref{fig:biasedgauge}). Further, we observe that the few extreme values of the gauge functional reached by the procedure performed in Fig.~\ref{fig:biasedgauge} are far from the main peak of central distribution in Fig.~\ref{fig:densityPlots} corresponding to $\beta=2.5$. We therefore conclude, that they don't play a major role in the ensemble average.

The average values of the Creutz ratio (vs. $R_{MCG}$ data) are marked by a ``$\star$'' symbol in Fig.~\ref{fig:densityPlots}, which give first estimates of the string tension. From the distributions it is clear that the gauge configurations with extremely high values of the gauge functional~(\ref{GaugeFunctR}) do not contribute to the average. 
Creutz ratios from larger Wilson loops show broader distributions due to larger statistical errors and give therefore less confidential values for the string tension than fitting the static potential. Therefore we have a look at the static potentials and fits thereof in the next sections, from which we can extract the string tension with high precision.

\subsection{Center Dominance and Precocious Linearity}
In this section, we compare the potentials extracted from original configurations with those which we get after center projections, especially with the ensemble averages over the local maxima of the gauge functional. To extract the physical parameters of the gluonic quark-antiquark string we fit the static potentials extracted from Wilson loops with a Cornell type of potential
\begin{equation}\label{eq:potfitallg}
V(R) = v_0 + \sigma R - \dfrac{c}{R}
\end{equation}
with three parameters, self energy $v_0$, string tension $\sigma$ and Coulomb coefficient $c$. 
For DLCG and EaMCG we observe the feature of precocious linearity; i.e.\ the fact that the Coulombic term is strongly suppressed. Therefore we choose $c=0$ and start the fit at some $R_i$.

\begin{figure}[h!]  
	\centering
	\includegraphics[scale=.75]{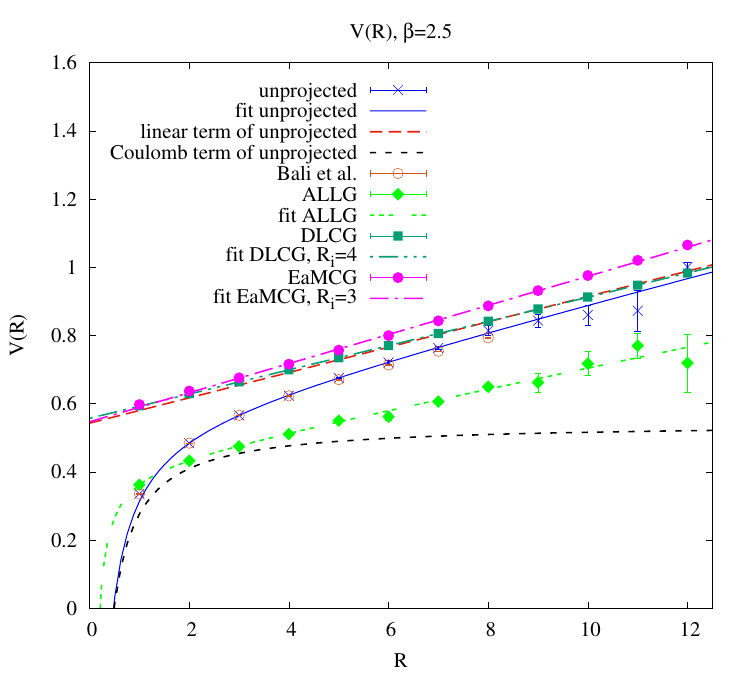}
	\caption{\small For $\beta=2.5$ on a $24^4$ lattice, potentials determined with different methods are shown and compared to Ref~\cite{Bali1994}. To ease comparisons, the potentials of EaMCG and DLCG are shifted by the self energy $v_0$ of the potential extracted from unprojected configurations.}
	\label{fig:fullshproj2}
\end{figure}
For $\beta=2.5$ we show in Fig. \ref{fig:fullshproj2} the potential values of unprojected configurations, the fit potential $V(R)$, its contributions $v_0-\dfrac{c}{R}$ and $v_0+\sigma R$ together with the data of Ref.~\cite{Bali1994}. We compare them with the potentials and their fits resulting from the three center projection methods, ALLG, DLCG and EaMCG. ALLG has a sizeable Coulomb contribution, differing in this respect from DLCG and EaMCG. Its potential is more difficult to resolve than for DLCG and EaMCG. ALLG has a smaller linear term than DLCG, where this term agrees nicely with the linear term of the unprojected configurations. Over the whole $R$ region the EaMCG values are very well resolvable and start at $R_i=3$ to predict a stable value for the string tension $\sigma$. It is interesting to observe in Fig. \ref{fig:fullshproj2} that, despite the linear term is larger than for DLCG and the unprojected configurations, the potential runs at large $R$ nicely parallel to the unprojected potential. When fitting the potential from unprojected configurations, the Coulomb factor is reasonably close to $\pi/12$, fixing the latter value in the fit changes the other fit parameters more or less within errors only, as can be seen from the numbers in Tab. \ref{table:potpar}.

\begin{figure}[h!]  
	\centering
	\includegraphics[width=0.47\linewidth]{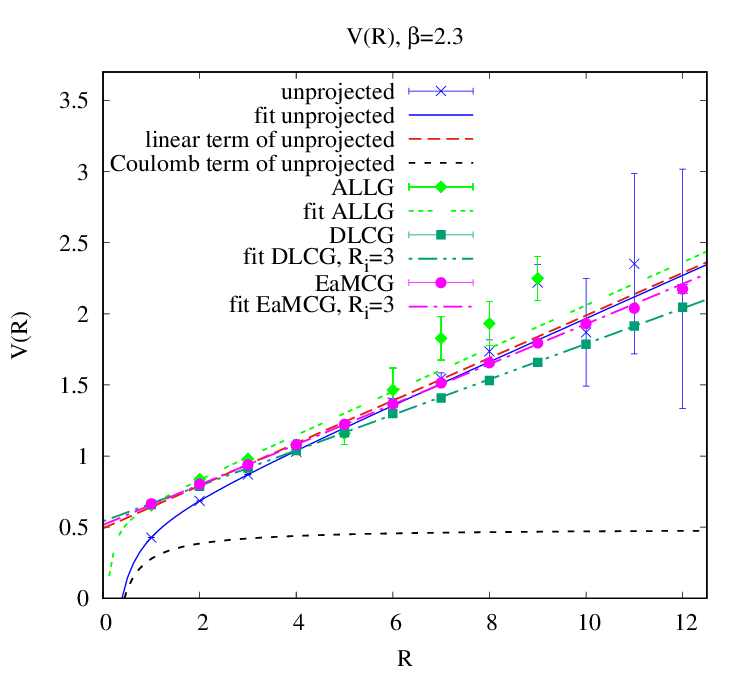}
	\includegraphics[width=0.47\linewidth]{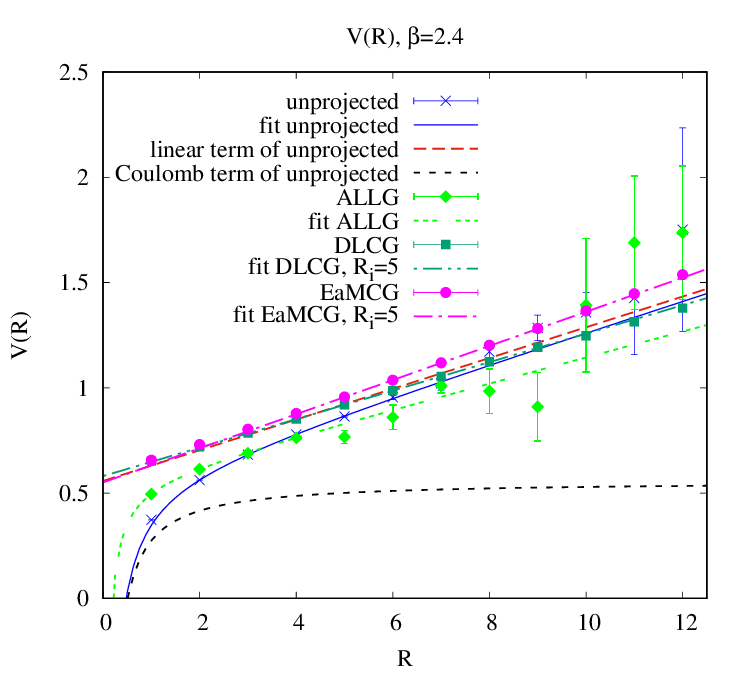}
	\includegraphics[width=0.47\linewidth]{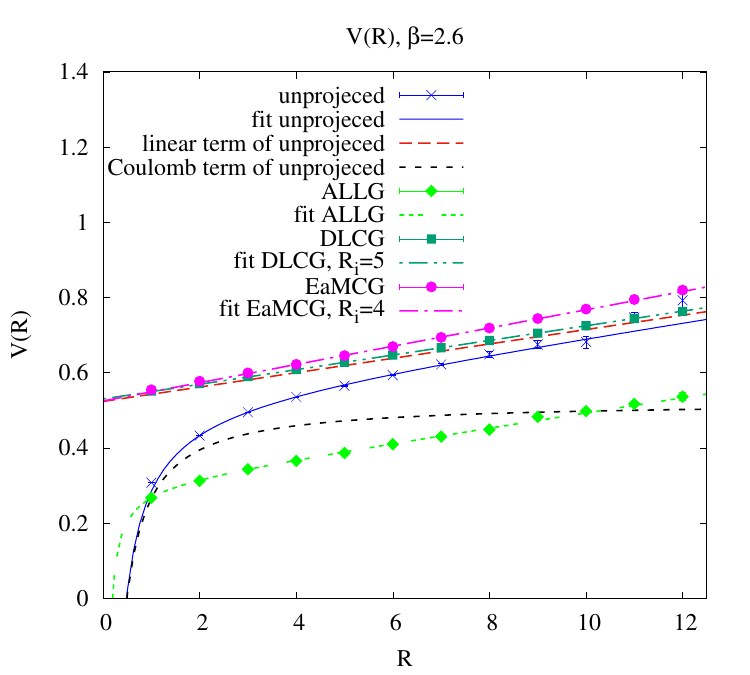}
	\includegraphics[width=0.47\linewidth]{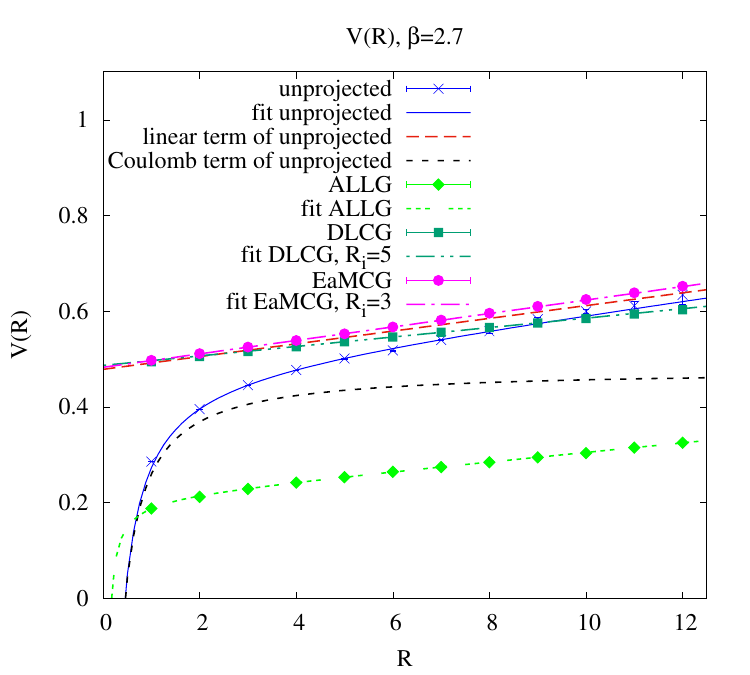}
	\caption{\small For $\beta\in\{2.3, 2.4, 2.6, 2.7\}$ on a $24^4$ lattice, the different methods to extract potentials and string tension are compared. For better visual comparison, the potentials of EaMCG and DLCG are shifted by the self energies of potentials extracted from original configurations.}
	\label{fig:fullshproj3}
\end{figure}

The parameters of the potentials are summarized in table \ref{table:potpar}. The indicated errors are the statistical errors only. The systematic errors are difficult to quantify, but they are at least of the same size as the statistical errors as we can see comparing the data.
\begin{table}[!htb]
	\centering	
	\begin{tabular}{l*{6}{c}r}	
		\hline
		\textbf{method} & $v_0$ & $\sigma$ & $c$  \\
		\hline
		$\beta=2.3$ \\
		\hline
		unprojected, $24^4$ & $0.492(8)$ & $0.1496(27)$ & $0.214(6)$ \\
		unprojected, $24^4$, fix $c$ & $0.5289(43)$ & $0.1431(21)$ & $\pi/12$  \\
		unprojected, $32^4$ & $0.495(5)$ & $0.1490(16)$ & $0.216(3)$ \\
		$Z_2$, ALLG &  $0.557(66)$ & $0.1508(193)$ & $0.053(48)$ \\
		$Z_2$, DLCG & $0.0496(18)$&$0.1247(6)$ & $-$ \\
		$Z_2$, EaMCG &  $0.0237(13)$ & $0.1412(4)$ & $-$\\
		\hline
		$\beta=2.4$ \\
		\hline
		unprojected, $24^4$ &$0.557(15)$ & $0.0730(25)$ & $0.284(20)$ \\
		unprojected, $24^4$, fix $c$ & $0.5472(41)$ & $0.0741(12)$ & $\pi/12$  \\
		unprojected, $32^4$ & $0.552(10)$ & $0.0728(17)$ & $0.271(13)$   \\
		$Z_2$, ALLG & $0.547(22)$ & $0.0608(56)$ & $0.112(16)$ \\
		$Z_2$, DLCG & $0.0233(17)$ & $0.0676(3)$ & $-$ \\
		$Z_2$, EaMCG  & $-0.0070(21)$ & $0.0811(4)$ & $-$ \\
		\hline
		$\beta=2.5$ \\
		\hline
		unprojected, $24^4$  &$0.543(5)$ & $0.0371(8)$ & $0.266(6)$  \\
		unprojected, $24^4$, fix $c$ & $0.5410(16)$ & $0.0373(5)$ & $\pi/12$   \\
		unprojected, $32^4$ &  $0.552(5)$ & $0.0350(8)$ & $0.276(6)$  \\
		$Z_2$, ALLG &  $0.413(3)$ & $0.0299(7)$ & $0.081(2)$  \\
		$Z_2$, DLCG &  $0.01423(9)$ & $0.03550(2)$ & $-$  \\
		$Z_2$, EaMCG &  $0.0040(14)$ & $0.0427(3)$ & $-$   \\
		\hline
		$\beta=2.6$ \\
		\hline
		unprojected, $24^4$ &$0.524(4)$ & $0.0191(6)$ & $0.258(5)$  \\
		unprojected, $24^4$, fix $c$  &$0.5256(12)$ & $0.0189(3)$ & $\pi/12$   \\
		unprojected, $32^4$  & $0.518(3)$ & $0.0200(5)$ & $0.251(4)$ \\
		$Z_2$, ALLG &  $0.302(2)$ & $0.0197(5)$ & $0.054(2)$  \\
		$Z_2$, DLCG &  $0.00655(11)$ & $0.01946(2)$ & $-$  \\
		$Z_2$, EaMCG &  $0.0013(8)$ & $0.0242(2)$ & $-$ \\
		\hline
		$\beta=2.7$ \\
		\hline
		unprojected, $24^4$  &$0.479(3)$ & $0.0133(5)$ & $0.219(5)$  \\
		unprojected, $24^4$, fix $c$ &  $0.4951(31)$ & $0.0118(7)$ & $\pi/12$   \\
		unprojected, $32^4$ & $0.489(2)$ & $0.0116(3)$ & $0.234(3)$  \\
		$Z_2$, ALLG &  $0.211(1)$ & $0.0097(2)$ & $0.034(2)$ \\
		$Z_2$, DLCG &  $0.00822(34)$ & $0.00988(6)$ & $-$  \\
		$Z_2$, EaMCG  & $0.0042(2)$ & $0.01406(6)$ & $-$  \\
		\hline
	\end{tabular}
	\caption{\small The parameters, self energy $v_0$, string tension $\sigma$, Coulomb factor $c$ of the potential fits (\ref{eq:potfitallg}) are compared for the various methods.}
	\label{table:potpar}
\end{table}

\subsection{Scaling behavior}
\begin{figure}[!htb]  
	\centering
	\includegraphics[scale=.7]{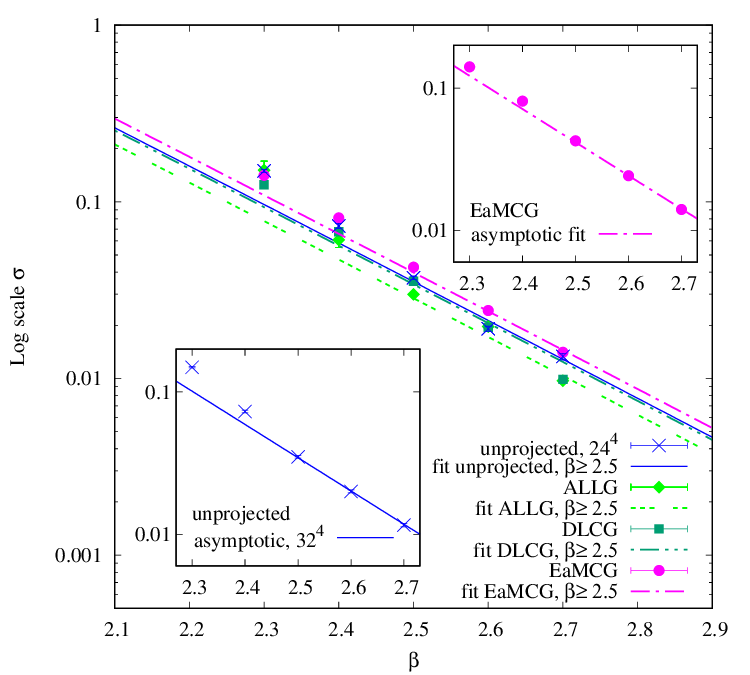}
	\caption{\small The $\beta$-dependence of the string tensions extracted from unprojected and center projected configurations is compared with the predictions in Eq.\eqref{eq:eq26prd} in the main figure. The  insets show fits to the asymptotic behavior $a^2\rightarrow 1 / {\Lambda_L^2} \exp({-\dfrac{6 \pi^2}{11}\beta})$. For each of the methods the lattice scale parameter $\Lambda_L$ is adjusted separately to the data with $\beta \geqslant 2.5$. The error bars are underestimated because we did not take (auto-)correlations and other systematic effects into account.}
	\label{fig:stringtension2}
\end{figure}

The string tension is expected to be proportional to the square of the lattice spacing $a$. For the SU(2) gauge group in the quenched approximation~\cite{rothe1992} asymptotic freedom predicts in two-loop order
\begin{equation}\label{eq:eq26prd}
a^2 = \dfrac{1}{\Lambda_L^2} (\dfrac{6 \pi^2 \beta}{11})^{\dfrac{102}{121}} \exp({-\dfrac{6 \pi^2}{11}\beta}),
\end{equation}
where $\Lambda_L$ is the standard lattice scale parameter. In Fig.~\ref{fig:stringtension2} we compare the string tensions extracted from unprojected configurations and the three methods of center projection, see Table~\ref{table:potpar}, with the asymptotic freedom prediction of Eq.~(\ref{eq:eq26prd}). The main diagram shows the results of our calculations on a $24^4$ lattice. In the insets we show fits to the asymptotic behavior $a^2\rightarrow 1 / {\Lambda_L^2} \exp({-\dfrac{6 \pi^2}{11}\beta})$, in the upper inset for EaMCG and in the lower inset for unprojected configurations on a $32^4$ lattice. All our obtained results, including EaMCG data, approach the expected asymptotic scaling with increasing $\beta$.

\section{Conclusion}
Center gauge methods suffer from the large differences between the center projected and unprojected field configurations. These differences appear especially in the region of center vortices. Unconditionally maximizing the center blind gauge functional~(\ref{GaugeFunctR}) underestimates the vortex density and therefore the string tension. This problem is related to an approximate linear relation between the gauge functional and the string tension as shown in Figs.~\ref{fig:biasedgauge} and \ref{fig:densityPlots}. This indicates that the request for a global maximization of the gauge functional is erroneous.

There were different suggestions to overcome this failure of the global maximization condition. A first idea~\cite{Faber_2001,Faber_2002} was to preselect a region of smooth gauge configurations where we maximize the gauge functional. With the three lowest eigenfunctions of the center blind lattice Laplacian in the adjoint representation one defines in ALLG a smooth gauge transformed configuration to start the maximization procedure by a gradient ascent to arrive at DLCG. In~\cite{Rudolf_Golubich_93027173,Golubich_2021,Rudolf_Golubich_68347507}  we suggested a completely different, but computationally rather involved approach, where we proposed to use global structures, non-trivial center regions, to guide the gauge fixing procedure.

The investigations in this article support the importance of the ensemble versus global maximization. This could possibly hint at a conditional maximisation prescription within the ensemble, requiring further investigations. Such conditional maximisation procedure is given by the aforementioned non-trivial center regions and proofed promising for single configurations.

An attempt to systematically address the issue of a heuristically defined gauge has been presented in the context of Coulomb gauge~\cite{Heinzl:2008bu}.

Recently, an improved method for computing the static quark-antiquark potential in lattice QCD was investigated in \cite{Hollwieser:2022doy, Hollwieser:2022pov, Hollwieser:2022bqp}, which is not based on Wilson loops, but formulated in terms of (temporal) {\it Laplace trial state correlators}, formed by eigenvector components of the covariant lattice Laplace operator. This approach seems very promising and an application to center vortex configurations as well as a generalization of the methods introduced in this article to $SU(3)$ is left for future investigations.  

\section*{Acknowledgements} The authors gratefully acknowledge the Vienna Scientific Cluster (VSC) for providing supercomputer resources. R.H. was supported by the MKW NRW under the funding code NW21-024-A. We thank Sedigheh Deldar for valuable discussions.

\bibliographystyle{utphys}
\bibliography{references}
\end{document}